\documentclass[pdflatex,sn-mathphys-num]{sn-jnl}

\usepackage{graphicx}%
\usepackage{multirow}%
\usepackage{amsmath,amssymb,amsfonts}%
\usepackage{amsthm}%
\usepackage{mathrsfs}%
\usepackage[title]{appendix}%
\usepackage{xcolor}%
\usepackage{textcomp}%
\usepackage{manyfoot}%
\usepackage{booktabs}%
\usepackage{algorithm}%
\usepackage{algorithmicx}%
\usepackage{algpseudocode}%
\usepackage{listings}%
\usepackage{array}
\usepackage{multirow}
\usepackage{lineno}
\usepackage{verbatim}
\usepackage{xcolor}

\usepackage[skip=5pt]{caption}
\usepackage{adjustbox}
\newcommand{\ours}{OMG-HD}
\begin{document}
\title[]{\ours{}: A High-Resolution AI Weather Model for End-to-End Forecasts from Observations}

\author[1]{\fnm{Pengcheng} \sur{Zhao}}
 
\author[1]{\fnm{Jiang} \sur{Bian}}
 
\author[1]{\fnm{Zekun} \sur{Ni}}

\author[1]{\fnm{Weixin} \sur{Jin}}

\author[1]{\fnm{Jonathan} \sur{Weyn}}
 
\author[1]{\fnm{Zuliang} \sur{Fang}}
 
\author[1]{\fnm{Siqi} \sur{Xiang}}
 
\author*[1]{\fnm{Haiyu} \sur{Dong}}\email{Haiyu.Dong@microsoft.com}
 
\author[1]{\fnm{Bin} \sur{Zhang}}

\author[1]{\fnm{Hongyu} \sur{Sun}}
 
\author[1]{\fnm{Kit} \sur{Thambiratnam}}
 
\author[1]{\fnm{Qi} \sur{Zhang}}
 
\affil[1]{\orgdiv{Microsoft Corporation}}

\abstract{ 
    In recent years, Artificial Intelligence Weather Prediction (AIWP) models have achieved performance comparable to, or even surpassing, traditional Numerical Weather Prediction (NWP) models by leveraging reanalysis data. However, a less-explored approach involves training AIWP models directly on observational data, enhancing computational efficiency and improving forecast accuracy by reducing the uncertainties introduced through data assimilation processes.
    In this study, we propose \ours{}, a novel AI-based regional high-resolution weather forecasting model designed to make predictions directly from observational data sources, including surface stations, radar, and satellite, thereby removing the need for operational data assimilation.
    Our evaluation shows that \ours{} outperforms both the European Centre for Medium-Range Weather Forecasts (ECMWF)'s high-resolution operational forecasting system, IFS-HRES, and the High-Resolution Rapid Refresh (HRRR) model at lead times of up to 12 hours across the contiguous United States (CONUS) region.
    We achieve up to a 13\% improvement on RMSE for 2-meter temperature, 17\% on 10-meter wind speed, 48\% on 2-meter specific humidity, and 32\% on surface pressure compared to HRRR. 
    Our method shows that it is possible to use AI-driven approaches for rapid weather predictions without relying on NWP-derived weather fields as model input. This is a promising step towards using observational data directly to make operational forecasts with AIWP models.
}

\keywords{Machine learning, Weather forecast, Earth observations}

\maketitle

\section{Introduction}\label{Introduction}

Over the last few decades, numerical weather prediction (NWP) has undergone what has been called a ``quiet revolution''~\cite{bauer2015quiet}, with steady improvements in forecast accuracy driven by advances in computing power, data assimilation, and physical parameterizations. In recent years, the introduction of artificial intelligence weather prediction (AIWP) models~\cite{pathak2022fourcastnet,chen2023fengwu,chen2023fuxi,bi2023accurate,lam2023learning,bodnar2024aurora} has provided an alternative approach, demonstrating superior globally-averaged accuracy compared to traditional NWP models while operating at a tiny fraction of their computational cost.

The availability of a rich set of reanalysis datasets, such as the European Centre for Medium-Range Weather Forecasts (ECMWF) Reanalysis v5 (ERA5)~\cite{hersbach2020era5}, makes the field of atmospheric prediction suitable for machine learning approaches. 
As convenient as reanalysis datasets may be, their basis in NWP means that they are an approximation to actual observations~\cite{eyre2022assimilation,gustafsson2018survey}. Biases in NWP forecasts can stem from incomplete representations of atmospheric physical processes, insufficient spatio-temporal resolution, and uncertainties in empirical parameterizations~\cite{laloyaux2022deep}. 
The data assimilation process, by which raw observational data are integrated into a gridded atmospheric state space, inherently results in information loss due to the limited resolution of the model grid and the approximations in the observation operators~\cite{valmassoi2023current,hu2023progress}. In fact, the state-of-the-art data assimilation process run at the ECMWF is only able to leverage about 5–10\% of the total satellite data volume~\cite{bauer2015quiet}. AIWP models, however, are not necessarily constrained by the assimilation methodology. Instead, they can learn latent representations of observational data, potentially addressing limitations inherent in NWP model physics.

The advantage of shifting away from NWP-generated data has motivated the development of End-to-End (E2E) models, whereby an AIWP model is trained directly on observational data, bypassing the need for intermediary NWP data. This approach offers the benefit of responsiveness, as E2E models can be updated rapidly in real-time as new observational data becomes available, eliminating the need for costly data assimilation processes and enabling a computationally efficient response to shifts in weather patterns. Additionally, E2E models may improve forecast accuracy by leveraging abundant and diverse observational data while avoiding the uncertainties inherent in NWP models, allowing them to better represent ground truth and uncover latent features that traditional NWP methods may miss.

A few E2E AIWP systems have already been developed \cite{fu2024lightweather,xiao2023fengwu,wu2023interpretable,vaughan2024aardvark,mcnally2024data,chen2023towards,sun2024fuxi,andrychowicz2023deep}, but so far none take full advantage of the benefits of this approach. For instance, \cite{wu2023interpretable} employs limited observation station data without diverse input such as satellite and radar data, which restricts model generalization; \cite{xiao2023fengwu,chen2023towards} do not fully decouple their models from NWP data as it is used as inputs for model training; \cite{vaughan2024aardvark,mcnally2024data,chen2023towards,sun2024fuxi} do not yet achieve performance surpassing traditional NWP. 
\begin{figure}[htbp]
    \centering
    \includegraphics[width=1\linewidth]{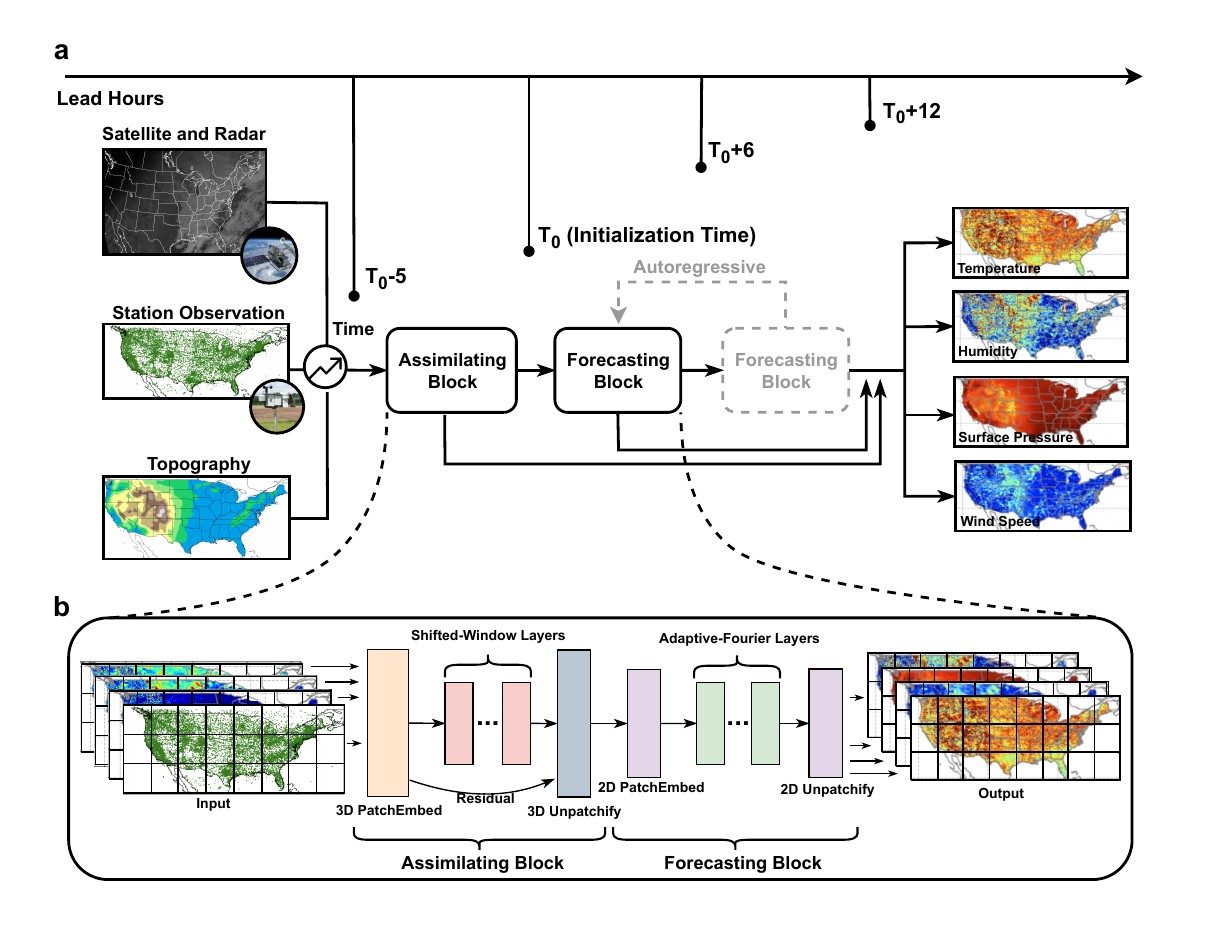}
    \caption{\textbf{The framework of \ours{}.} \textbf{a}, The forecasting pipeline with an annotated timeline. At the initial time $T_0$, observations (satellite, radar, and station measurements) and the topography from the preceding 6 hours ([$T_0-5$,$T_0$]) are passed through the Assimilating Block to produce an initial state, which is subsequently used by the Forecasting Block to predict the following 6 hours ([$T_0+1$,$T_0+6$]). The forecast can be fed back into the Forecasting Block in an autoregressive fashion to produce temperature, humidity, pressure, and wind predictions for the next 6 hours ([$T_0+7$,$T_0+12$]), and so on. \textbf{b}, Detailed architecture of the Assimilating Block and the Forecasting Block.}
    \label{fig:framework}
\end{figure}
We introduce \ours{}, an \underline{O}bservational \underline{M}eteorological data \underline{G}uided system for \underline{H}igh-\underline{D}efinition regional weather forecasting. \ours{} achieves the advantages of the E2E approach for operational forecasting by directly using raw observations as input, enabling operational responsiveness and accuracy that surpasses the High-Resolution Rapid Refresh (HRRR) model. Our main contributions are summarized as follows:
\begin{itemize}
    \item \textbf{Direct learning and inference from raw observational data}: \ours{} learns to make forecasts only from observational data, reducing the propagation of potential inaccuracies typically introduced by NWP-derived input data. 
    
    \item \textbf{Responsive and efficient real-time inference}: Using the latest observational data, \ours{} enables rapid adaptation to weather changes, avoiding the need to wait for the lengthy process of traditional data assimilation.
    
    \item \textbf{Accurate short-term regional forecasting}: Within a 12-hour forecast window, \ours{} consistently surpasses established models such as IFS-HRES and HRRR across the continental US, establishing a new benchmark in regional AIWP. Ablation studies further highlight the robustness of \ours{} to missing input data.
    
    \item \textbf{A launchpad for further research in E2E forecasting}: \ours{} introduces a novel paradigm for transforming raw data into forecasts with a simple yet effective model structure, paving the way forward for further experimentation to push the accuracy limits or build a forecast system completely independent of NWP.
\end{itemize}

\section{\ours{}}\label{overview}

The \ours{} model is specifically designed to process raw observational data from multiple sources for data-driven forecasting. The model framework and architecture are shown in Fig.~\ref{fig:framework}. Its architecture comprises two key components: the Assimilating Block and the Forecasting Block.

At the forecast issue time $T_0$, multi-source observations from the preceding six hours ($T_0 - 5 \sim T_0$) are ingested into the Assimilating Block through specially designed input layers that process diverse observation types, including in-situ and remote-sensing data. The Assimilating Block transforms such data into a gridded representation, bridging the gap between wide-ranging, heterogeneous observations and a more comprehensive and structured modeling of the current condition.
Details of the channels in the gridded output are introduced in Section~\ref{method}. A portion of these channels is directly supervised in the loss function, while others capture latent space information, representing features other than surface variables, such as upper air conditions. Subsequently, the Forecasting Block leverages this structured representation to generate predictions for the upcoming six hours ($T_0 + 1 \sim T_0 + 6$) and then iteratively forecasts the subsequent six hours ($T_0 + 7 \sim T_0 + 12$) in an autoregressive fashion.

For the Assimilating Block, we choose a multi-scale Transformer~\cite{liu2021swin} with Shifted-Window (Swin) Layers, which has been demonstrated in multiple applications to scale well to large spatio-temporal forecasting tasks including weather forecasting \cite{bi2023accurate,bodnar2024aurora}. 
Using the Adaptive-Fourier Layers for spatial mixing in the Forecasting Block \cite{guibas2021adaptive,pathak2022fourcastnet} significantly reduces computational complexity while managing higher input resolutions. The two blocks are trained end-to-end under a unified loss function, aligning the latent space representation generated by the Assimilating Block with the expectations from the Forecasting Block. More details on the model architecture and training process are provided in Section~\ref{model_arch}.

In this paper, \ours{} is trained on data over the contiguous United States (CONUS), leveraging rich data provided by over 18,000 surface stations, multi-channel satellite data and radar data, as detailed in Section~\ref{method}. 
During the training phase, the labels are derived from the Real-Time Mesoscale Analysis (RTMA)~\cite{de2011real}, which is a dataset specifically designed to align more closely with observations than traditional data assimilation schemes~\cite{morris2020quality}. 
Although RTMA relies on NWP, it vastly increases the number and diversity of targets in contrast to labels such as station observations (see Section~\ref{data_all}), without compromising our goal of producing operational forecasts based solely on observational inputs. 

\section{Results}\label{results}

In following sections, we conduct a comparative analysis of \ours{} against NWP-based forecasts, including IFS-HRES~\cite{holm2016new} (ECMWF), HRRR~\cite{dowell2022high,james2022high}, and GFS~\cite{zhou2019development}.
We evaluate \ours{} and the baselines on station- and grid-based targets for 12-hour forecasts over land points only. Our focus is on 2-meter temperature (T), 10-meter wind speed (WS), 2-meter specific humidity (Q), and surface pressure (SP), variables that directly measure impactful weather phenomena \cite{bouallegue2023statistical,bewoor2021artificial}. Future work could also include forecasts of other important variables such as precipitation, cloud cover, and wind gusts.

\subsection{Average forecast accuracy}\label{general}

\begin{figure}[htbp]
    \centering
\includegraphics[width=1\linewidth]{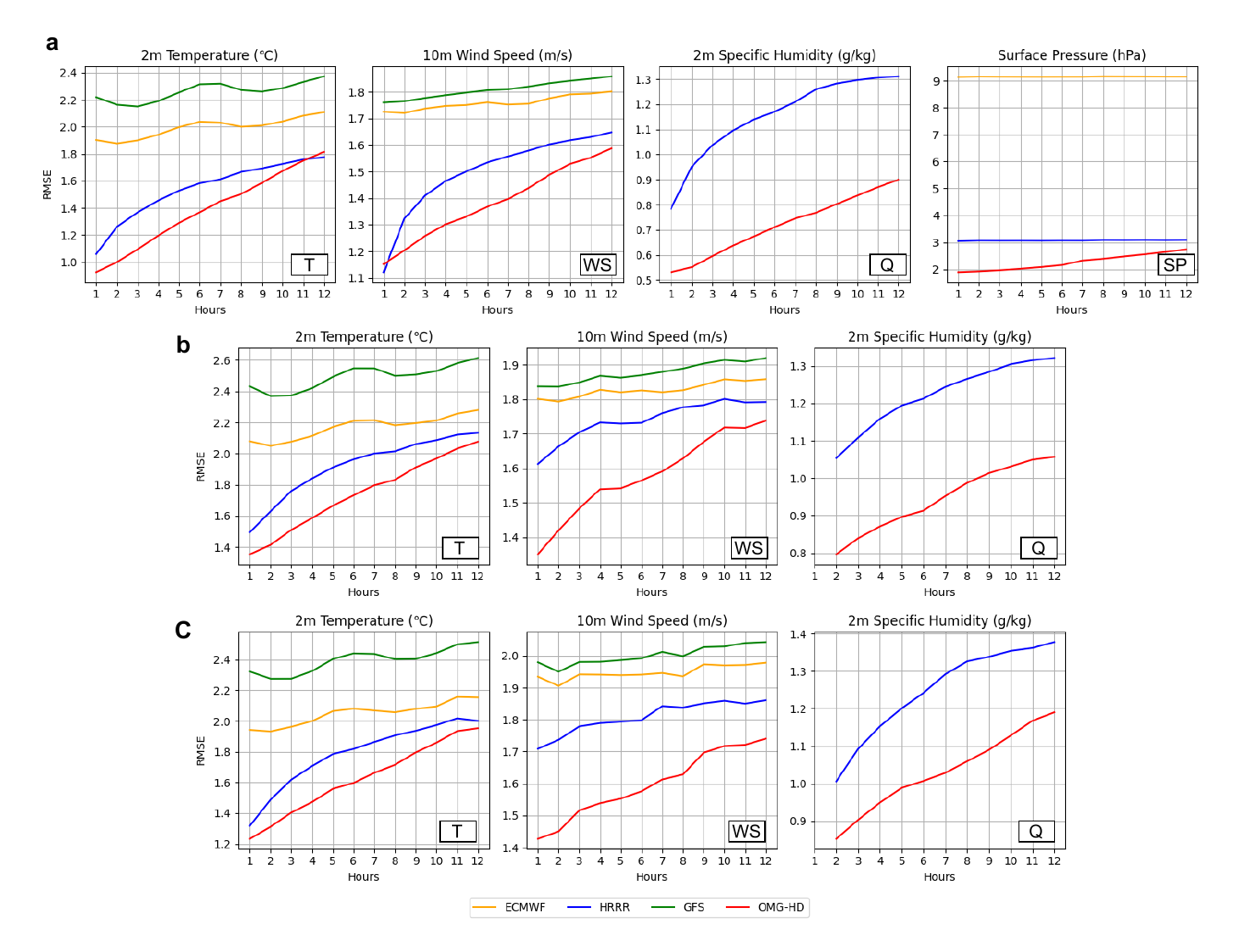}
    \caption{\textbf{\ours{} achieves lower forecasting error than baselines across varying lead hours.} \textbf{a}, RMSE for temperature, wind speed, specific humidity, and surface pressure verified against the RTMA dataset. \textbf{b}, RMSE for temperature, wind speed, and specific humidity verified against station observations. \textbf{c}, As in \textbf{b} but instead evaluated only on the hold-out set of observations (see Section~\ref{ablation}). The absence of certain curves indicates some variables are unavailable in certain model evaluations.
}
    \label{fig:OMG-HD+_RTMA}
\end{figure}

Fig.~\ref{fig:OMG-HD+_RTMA}a shows the root mean squared error (RMSE) of the models' forecasts as a function of lead time, measured against the RTMA as truth. \ours{} consistently exhibits superior performance relative to the baseline models across all four variables, particularly for Q and SP for nearly all lead times. The performance gain averaged over lead times compared to HRRR is 13\%, 17\%, 48\%, and 32\% on T, WS, Q and SP, respectively. Note that this difference is exaggerated by the original resolution of the models, particularly regarding SP, which is strongly modulated by the definition of the model topography. Despite being challenging, \ours{} still learns the localized RTMA SP patterns effectively, even from sparse observational data. Because OMG-HD lacks boundary conditions and an explicit loss targeting the upper-air atmospheric state, its forecast accuracy diminishes over time. This informs our decision to measure the model for the first 12 hours only.

\begin{figure}[tbp]
    \centering
    \includegraphics[width=1\linewidth]{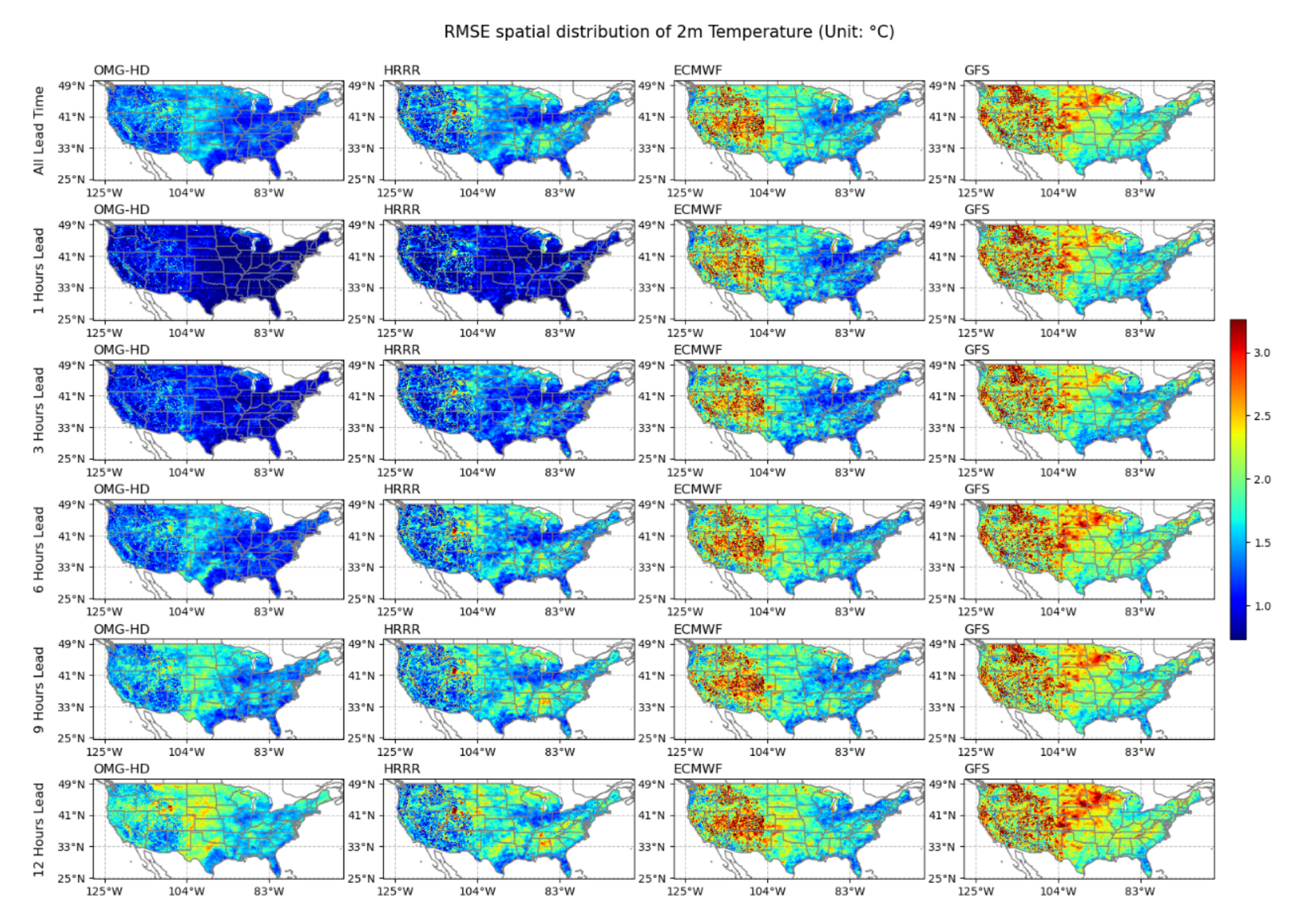}
    \caption{\textbf{\ours{} provides accurate, consistent temperature forecasts throughout the CONUS region.} Spatial distribution of temperature RMSE evaluated against the RTMA dataset for \ours{} and the baseline HRRR, ECMWF, and GFS models, as labeled. Average metrics are in the top row, while metrics for specific lead times of 1, 3, 6, 9, and 12 hours are shown in subsequent rows.
    }
    \label{fig:temp_vis}
\end{figure}

The evaluation on station observations is illustrated in Fig.~\ref{fig:OMG-HD+_RTMA}b. 
While the patterns align broadly with those in the evaluation against RTMA, the performance gap is more noticeable for variables T, WS, and Q, where \ours{} distinctly outperforms the baseline models. 
The prediction of surface pressure is significantly influenced by terrain. Considering the limited number of stations reporting surface pressure (fewer than 2,000) and their sparse, uneven spatial distribution, evaluating baseline models using station observations for surface pressure would be neither fair nor representative. Therefore, we have decided not to include surface pressure in the station observation evaluation.
Overall, \ours{} maintains an impressive lead over the NWP baselines for 12-hour predictions. 

\begin{figure}[tbp]
    \centering
    \includegraphics[width=1\linewidth]{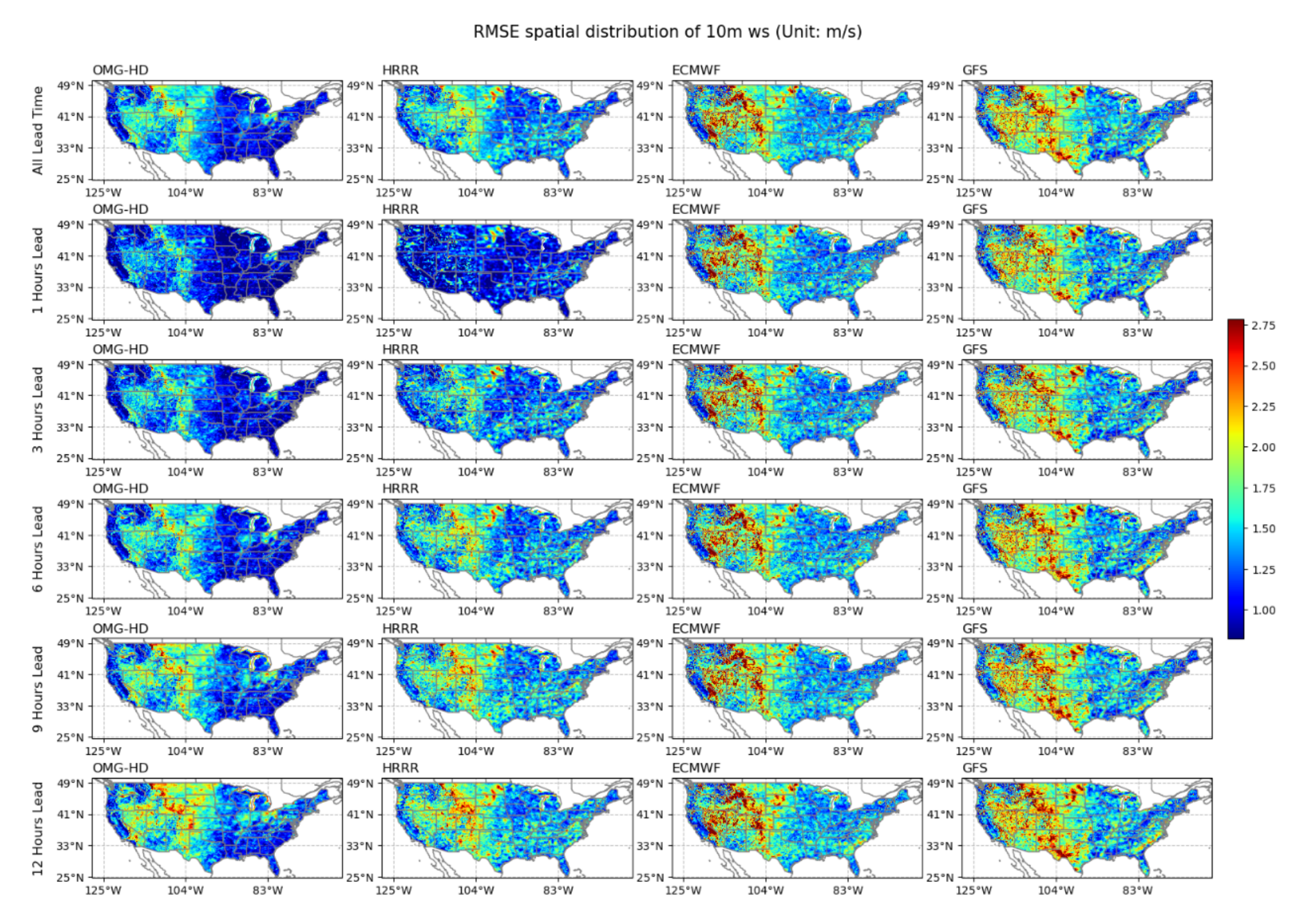}
    \caption{\textbf{\ours{} provides accurate, consistent wind speed forecasts throughout the CONUS region.} As in Fig.~\ref{fig:temp_vis} but for forecasts of wind speed.  
    }
    \label{fig:wind_vis}
\end{figure}

To better understand spatial variability in prediction quality, we also show the spatial distribution of the errors against RTMA for T in Fig.~\ref{fig:temp_vis} and WS in Fig.~\ref{fig:wind_vis} (see Section~\ref{app:additional} for the Q and SP results). Unsurprisingly, the coarser ECMWF and GFS models show large discrepancies in mountainous regions where the terrain accounts for much of the difference, but on the other hand they also develop significant errors over Central and Eastern US, where the orography is largely flat.
\ours{} exhibits the most spatially-consistent error patterns among all the models, with slightly higher RMSE only in the Great Plains region by hour 12.

\begin{figure}[htbp]
    \centering
    \includegraphics[width=1\linewidth]{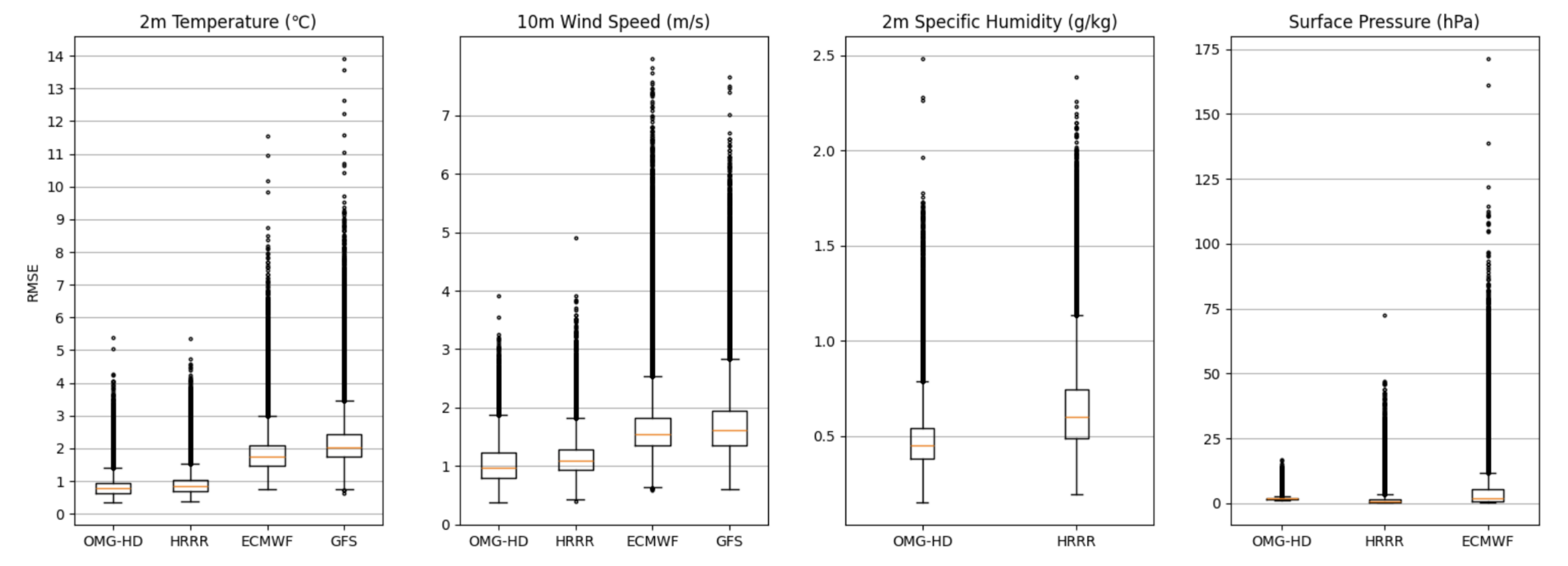}
    \caption{\textbf{The initial state produced by \ours{} is more accurate than that of the baselines.} Shown are box-and-whisker plots of RMSE values for all forecast grid points compared to RTMA. \ours{} has the lowest average and inter-quartile range of RMSE across all variables, and the fewest outliers (points).
    }
    \label{fig:da}
\end{figure}

\subsection{Evaluation of the embedded assimilation state} 

As illustrated in Fig.~\ref{fig:framework}, the Assimilating Block integrates all input data from the past 6 hours to form the initial state at $T_0$, which is later used for forecasting. 
Hence the \ours{} Assimilating Block generates a prior, similar to the traditional data assimilation process. 
Given the importance of the initial conditions for forecasting accuracy \cite{lorenz1969predictability}, 
we compare the quality of \ours{}'s initial state with those of the baseline models, evaluated against the RTMA labels in Fig.~\ref{fig:da}. 
The results largely align with the overall forecasting performance shown in Fig.~\ref{fig:OMG-HD+_RTMA}: \ours{} consistently has the lowest mean RMSE values (red lines) across all four variables.
Additionally, \ours{} has fewer outliers. 

\subsection{Ablation tests}\label{ablation}

The following tests in this section further demonstrate the robustness of the \ours{} framework.

\subsubsection{Testing with hold-out stations}\label{case_1}

In this test, we use a hold-out mechanism \cite{andrychowicz2023deep} to isolate specific stations from the training dataset, reserving them as hold-out stations to test how well the model predicts unseen data. This helps ensure the model does not overfit on the training stations. The detailed process of hold-out mechanism is presented in Section~\ref{data_all}. As shown by the forecast error on hold-out stations in Fig.~\ref{fig:OMG-HD+_RTMA}c, \ours{} model maintains superior performance compared to standard baseline models even on this set of unseen data. 
This result substantiates the model's generalization potential across different spatial locations, both with and without station observations present.

\subsubsection{Inference with masked input station observations}\label{case_2}
In operational settings, observational input may be unavailable due to equipment malfunctions, issues in data ingestion, or other unforeseen circumstances. 
It is therefore important to evaluate the ability of a forecasting system like \ours{} to handle missing data issues gracefully, with minimal loss in prediction accuracy \cite{alley2019advances,schmude2024prithvi}.

\begin{figure}[htbp]
    \centering
\includegraphics[width=1\linewidth]{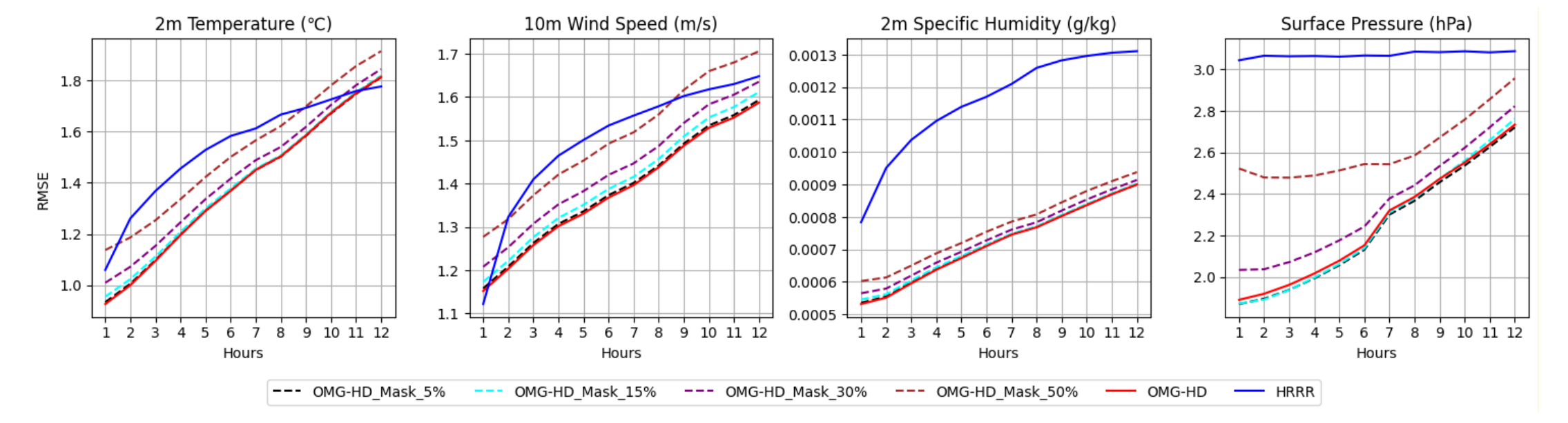}
    \caption{\textbf{Reducing the number of input stations induces minor performance drop for \ours{}.} Forecast RMSE as a function of lead time for T, WS, Q, and SP. Dashed curves represent \ours{} with 5\%, 15\%, 30\%, and 50\% masked station observations at inference time, in comparison with \ours{} without any input masking (solid red curves) and HRRR (solid blue curves). 
}
    \label{fig:mask_OMG-HD+}
\end{figure}
In reality, situations are far more complex and dynamic, making it difficult to completely reproduce them through simulations. Here, we adopt a commonly used method~\cite{gromov2023deconstructing,lad2024remarkable} to simulate real-world scenarios and evaluate the robustness of the model. 
To investigate the influence of missing data on our model, we randomly vary the masking ratio applied to input stations during the inference stage, adjusting it from 5\%, 15\%, 30\%, up to 50\% of all the input stations.
Fig.~\ref{fig:mask_OMG-HD+} shows the result of this data-denial experiment compared to the HRRR baseline. 
The accuracy of the forecasts exhibits only a relatively minor decline when more station data inputs are progressively removed.
With the exception of lead hours 1-2 and 9-12 for T and WS, \ours{} continues to outperform the HRRR model for all weather variables. In summary, \ours{} shows good robustness to missing observation data - an important quality, given imperfect input data in an operational setting.


\section{Case studies}\label{cases}

To evaluate the performance of \ours{} in impactful weather events, we've conducted a set of case studies.

\subsection{Northern plains winter storm in March 2023}\label{case_blizzard}

\begin{figure}[htbp]
    \centering
    \includegraphics[width=1\linewidth]{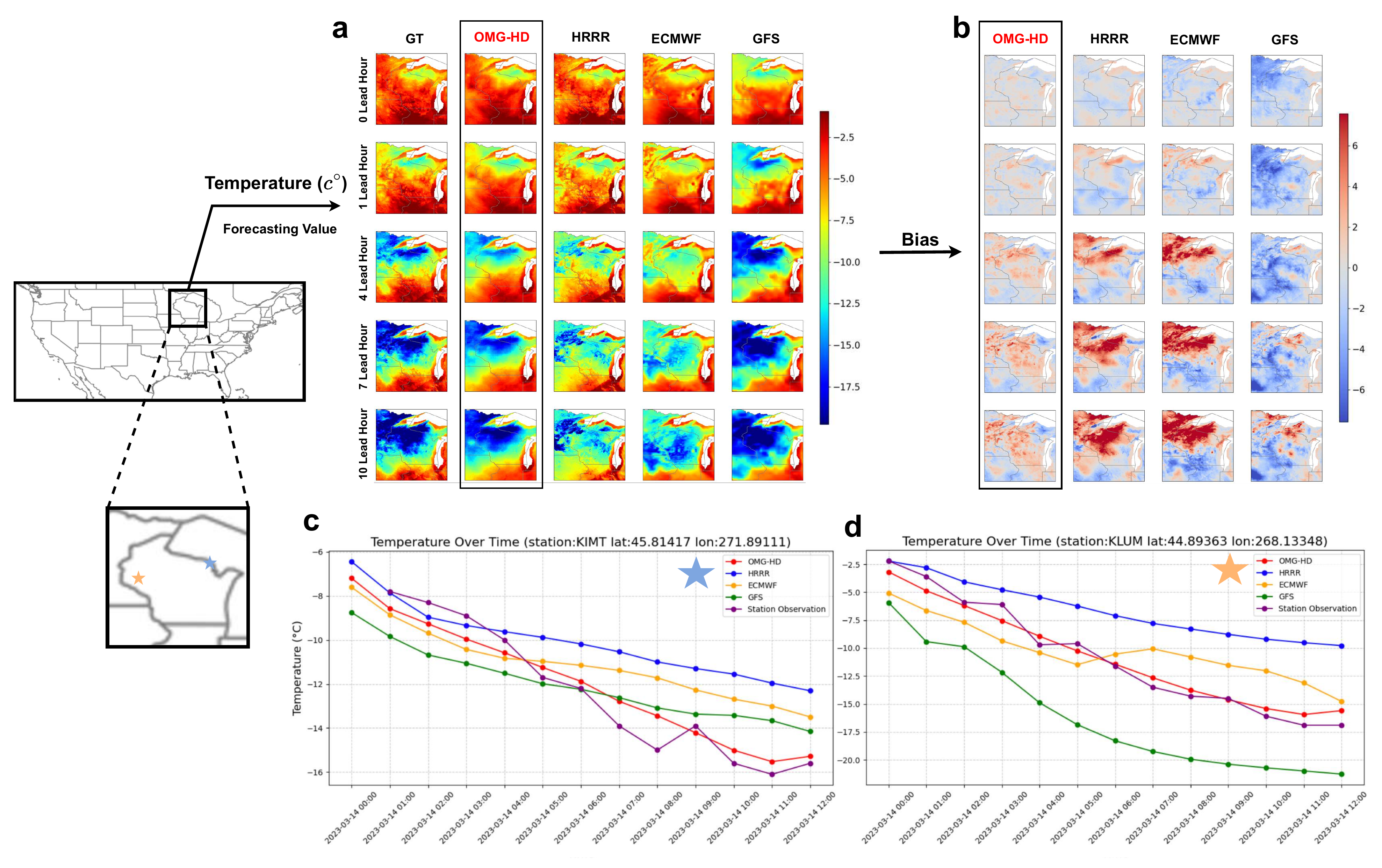}
    \caption{\textbf{\ours{} better predicts the temperature drop in a winter storm than the baselines.} A winter storm affected the Great Lakes region, USA, on March 14, 2023. \textbf{a} and \textbf{b}, Forecast values and their biases from ground truth over the selected area, where the columns correspond to different models, and rows show different lead times. \ours{} consistently exhibits the lowest bias across all lead times. \textbf{c} and \textbf{d}, Comparison of temperature forecasts from different models with actual station observations, on two sample stations. \ours{} closely aligns both station observations.
    }
    \label{fig:blizzard}
\end{figure}

\begin{figure}[htbp]
    \centering
    \includegraphics[width=1\linewidth]{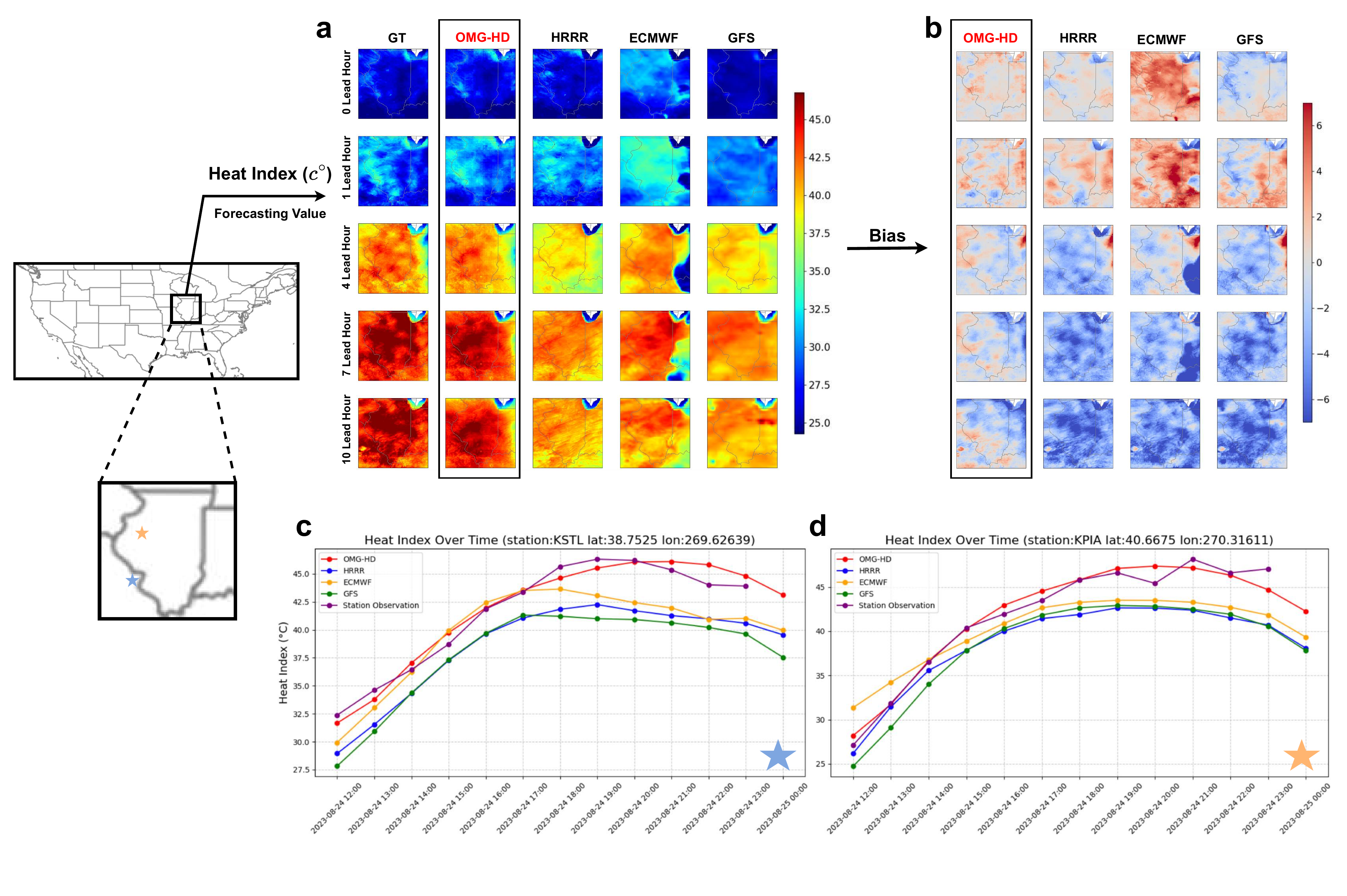}
    \caption{\textbf{\ours{} excels in predicting heatwave development by effectively capturing heat index increases.} Comparison of prediction accuracy for different models during the heatwave event of August 24, 2023 around Chicago, USA. \textbf{a} and \textbf{b}, Forecast values and their biases from ground truth over the selected area, where the columns correspond to different models, and rows show different lead times. \ours{} consistently exhibits the lowest bias across all lead times. \ours{} produces the best prediction of heat index compared to baseline models. \textbf{c} and \textbf{d}, Comparison of heat index forecasts from different models with actual station observations, on two sample stations. \ours{} closely aligns both station observations.}
    \label{fig:heatwave}
\end{figure}

In the first case, we examine a winter storm event which affected the Great Lakes region. The winter storm initially progressed across the Western US as an atmospheric river. As it moved across the northern US, it brought a series of extreme conditions including drastic temperature drops. Our forecast, issued at 0 UTC on 14 March 2023, captures the cold spell as it intensifies over Minnesota, Wisconsin, and Michigan, resulting in the distinct and sudden temperature drops shown in Fig.~\ref{fig:blizzard}. We compare temperature forecasts (including initial conditions) produced by \ours{} and various baseline models against the RTMA label. \ours{} produces predictions closely resembling the RTMA labels, exhibiting the lowest bias across all lead times (in Fig.~\ref{fig:blizzard}b). Here, we define the bias as the error between the forecast (F) and ground truth (GT) label, i.e., $\text{Bias} = \text{F} - \text{GT}$ . While the performance gap between \ours{} and HRRR is minimal at lead time zero, \ours{} gradually outperforms HRRR as the lead time increases. As shown in Fig.~\ref{fig:blizzard}a, \ours{} effectively captures the pattern of a rapid temperature drop in the precise area of interest. In contrast, HRRR and IFS-HRES (denoted as ECMWF in the figures) under-estimate the temperature drop. Meanwhile, the GFS forecast starts too cold and continues to over-estimate the temperature drop, reflected as larger, darker blue regions. Additionally, we selected two airport stations from METAR~\cite{lui2014complete} within the target area to examine in detail, as shown in Fig.~\ref{fig:blizzard}c and Fig.~\ref{fig:blizzard}d. Forecasts generated by \ours{} exhibit the highest accuracy, more closely aligning with station observations than those of other baseline models. Overall, in this case study, \ours{} demonstrates superior forecasting capability, accurately capturing abrupt temperature drop patterns associated with impactful events such as this winter storm.

\subsection{Chicago heatwave in August 2023}\label{case_heatwave}

In the second case, we investigate a heatwave affecting the Chicago area on August 24, 2023. An expansive ridge of high pressure over the central US led to several days of record-breaking temperatures in northern Illinois and northwest Indiana during mid-August. On August 24, a brief yet intense heat incident south of Chicago resulted in dangerously high temperatures. While the temperature forecast by \ours{} was slightly better than baselines, our model is much more successful in capturing the unusually high humidity (as shown in Fig.~\ref{fig:case_t_q}c), which had a significant impact on the elevated ``feels-like'' temperature. To capture this, we evaluate the performance of the models' forecasts of heat index, which better reflects the risk of heat stress.
Fig.~\ref{fig:heatwave}a and Fig.~\ref{fig:heatwave}b shows the forecasts and biases for this event. By lead hour 10, all models except \ours{} fail to capture the increase in heat index, mostly because they under-estimate the humidity levels (as shown in Fig.~\ref{fig:case_t_q}c).
While \ours{} still slightly under-estimates heat index near the Chicago metro area, it has much less bias in the forecast for this event compared to the baselines. In Fig.~\ref{fig:heatwave}c and Fig.~\ref{fig:heatwave}d, we selected two airport stations within the target region for closer examination. Among all models, \ours{} achieves the smallest gap between forecast values and station observations.

The case studies highlight the ability of \ours{} to deliver accurate and reliable forecasts for significant weather events, such as the aforementioned winter storm and heatwave, with implications for both safety and resource management. In the winter storm case, \ours{} effectively captured localized temperature drops, helping prepare for extreme winter conditions. For the heatwave, it accurately forecasted heat index levels, which are critical for addressing heat-related health risks. 

\section{Discussion}\label{discussion}

In this paper, we present OMG-HD, a data-driven kilometer-scale forecasting system that predicts directly from raw observations, including station, radar, and satellite data. Our evaluation of 12-hour forecasts over the CONUS region shows that it outperforms NWP models including HRRR, IFS-HRES, and GFS, across key variables including temperature, wind speed, specific humidity, and surface pressure, even with incomplete inputs. Case studies further highlight OMG-HD's ability to accurately capture extreme weather events.


Compared to recent AIWP models, OMG-HD exemplifies the advantages of the end-to-end paradigm. By directly learning from raw observational data, it minimizes information loss and improves the upper limit of forecasting accuracy. This may be one reason why the model demonstrates superior short-term performance over NWPs. Integrating the entire process also simplifies workflows, reducing the need for manual processing and consequently, the complexity of the whole system. Moreover, it eliminates the latency associated with generating analysis data, enabling rapid response to evolving weather variations.



Despite the promising accuracy, there are several directions worth exploring. First, OMG-HD currently uses RTMA as labels, which are limited to surface variables over the CONUS region and have small discrepancy from raw observations. 
This constraint restricts its broader application, as RTMA restricts the training process to focus only on the CONUS region, making it uncertain whether the resulting model can generalize to different regions, particularly those located in different latitude ranges or with differing surface conditions, such as those over the ocean. Moreover, without a global context, the model is unable to produce skillful forecasts beyond a few lead hours in absence of boundary conditions.
Second, the input data are still limited (e.g., only four satellite channels are employed), making it insufficient for fully capturing 3-dimensional atmospheric conditions. To address these issues, our next goal is to collect more comprehensive data and develop a “label system” to generate reliable training labels. With more input data and improved labels, we expect OMG-HD to achieve global coverage, predict additional variables, and extend to longer lead times.



The end-to-end design in \ours{} exhibits a promising future due to their responsiveness and short-term accuracy, suggesting significant potential to benefit the daily needs of users and support informed decision-making. By leveraging advances in data-driven methodologies, such approaches offer significant potential to enhance traditional weather forecasting paradigms. Similar advancements in end-to-end AI approaches have driven progress across various domains, including natural language processing and computer vision, showcasing their versatility and transformative capabilities~\cite{bontempi2024end, ates2022end,liao2024maptrv2}. We believe that such innovations hold similar promises for weather forecasting, enabling faster, more efficient, and highly adaptable systems that align with the increasing complexity of global weather patterns.

\bibliography{main}
\newpage
\appendix

\section{Methods}\label{method}

\subsection{Data}\label{data_all}

\begin{figure}[htbp]
    \centering
    \includegraphics[width=0.8\linewidth]{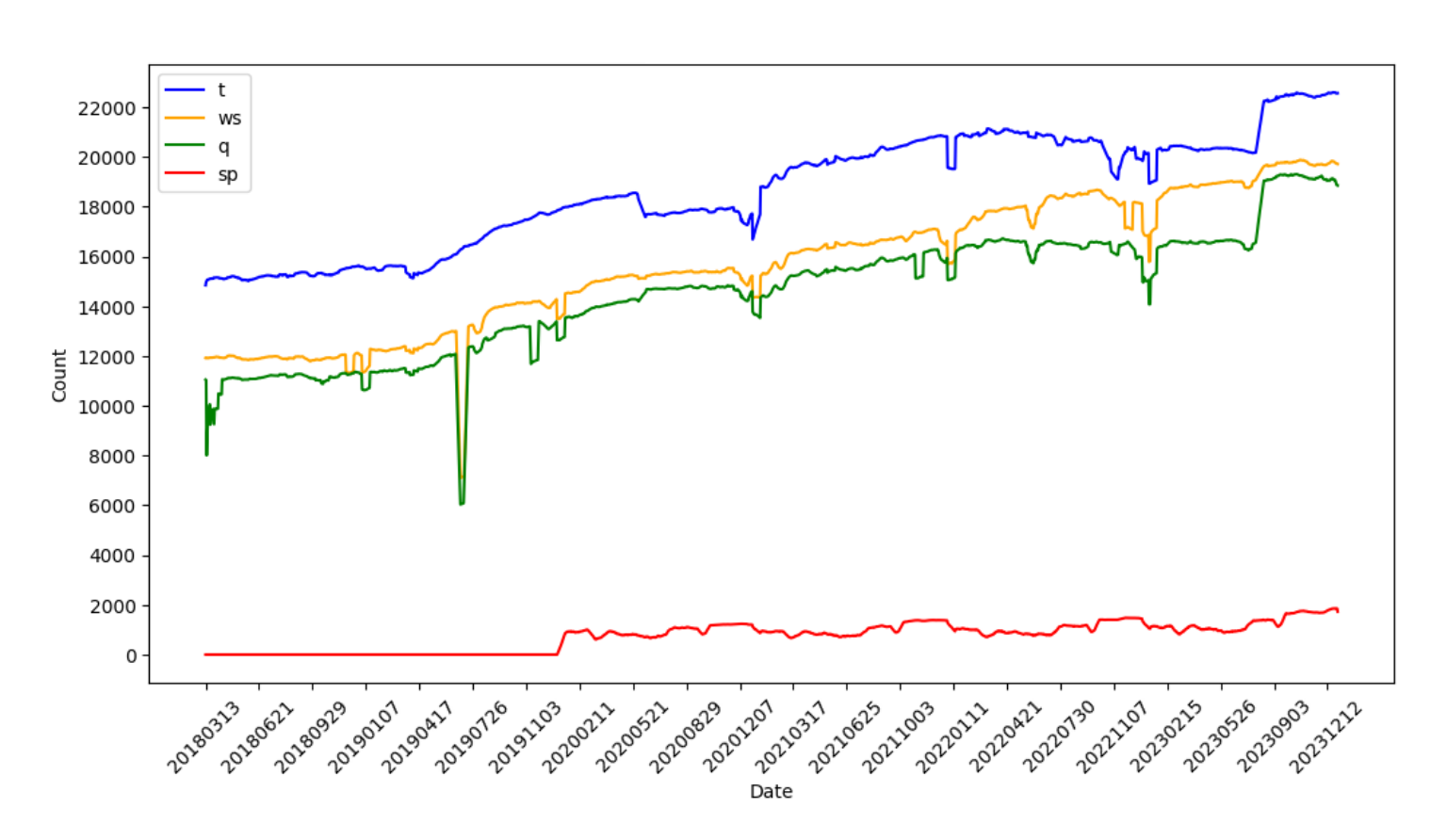}
    \caption{\textbf{Number of stations available for each variable over time.} Temporal distribution of available stations spanning from March 13, 2018, to December 12, 2023.}
    \label{fig:distribution}
\end{figure}
\begin{figure}[htbp]
    \centering
    \includegraphics[width=0.7\linewidth]{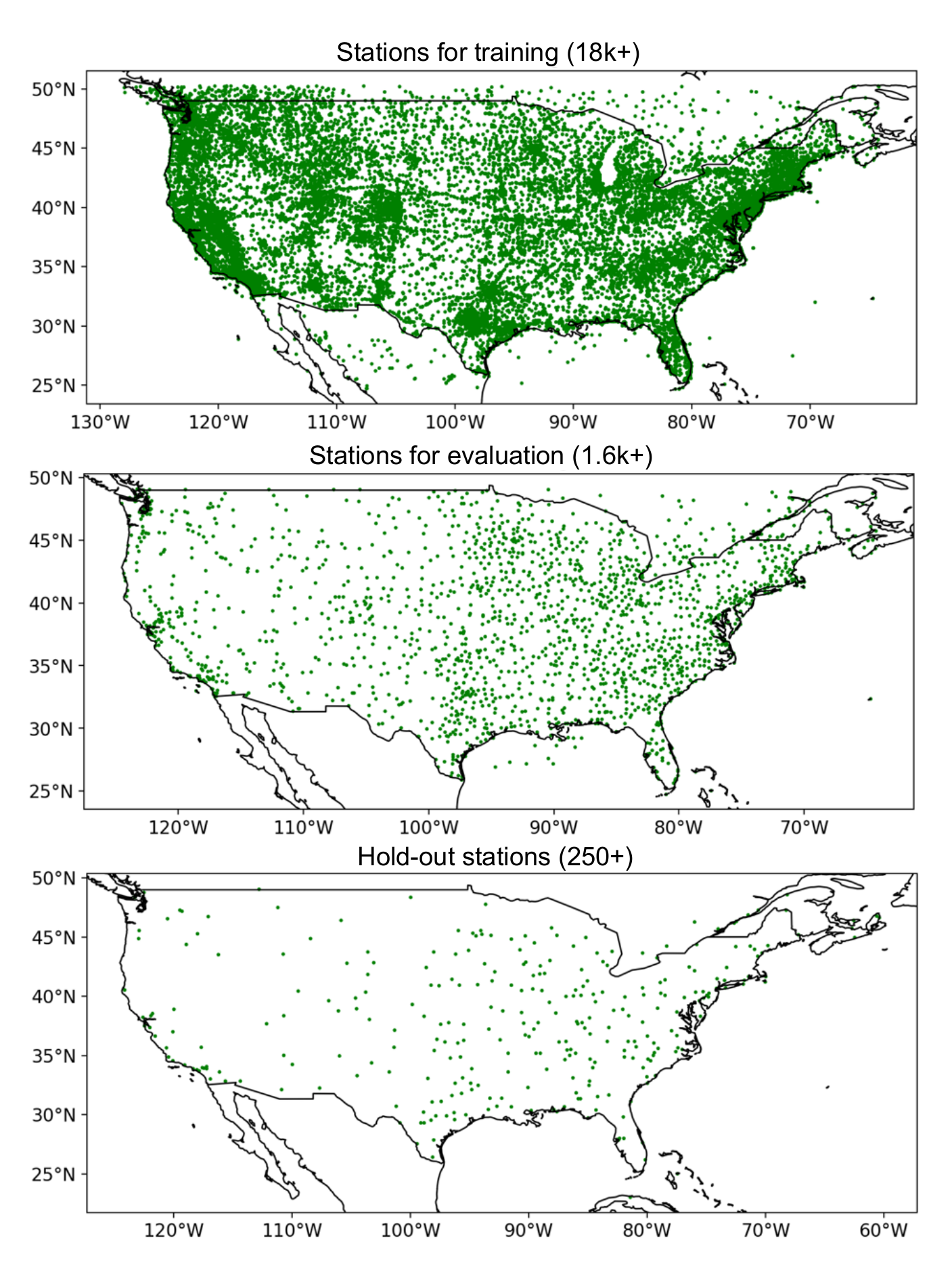}
    \caption{\textbf{Spatial distribution of stations for training, evaluation and hold-out purposes.} The panels from top to down illustrate: the spatial distribution of ground-based weather stations in the training phase, in evaluations, and the few hold-out stations in the training phase.} 
    \label{fig:train_evaluate_stations}
\end{figure}

In this study, we use  diverse data sources over the CONUS region, which is bounded between latitudes 24.70° to 50.25° north and longitudes 64.00° to 128.00° west, to train our model. First, we use weather stations derived from the WeatherReal-Synoptic dataset~\cite{jin2024weatherreal} and the Integrated Surface Database (ISD)~\cite{smith2011integrated}. The number of reporting stations over time for the four surface variables used (temperature, wind speed, specific humidity, and surface pressure) are illustrated in Fig.~\ref{fig:distribution}. The number of stations reporting T, WS, and Q varies from about 10,000 to 22,000 over the years 2018-2023, while significantly fewer stations (about 2,000 at most) report surface pressure (SP). 
The distribution of weather stations is depicted in Fig.~\ref{fig:train_evaluate_stations}. Following major population centers, stations are densely distributed along the east and west coasts, while being more sparse over the mountainous West and the Great Plains.
Next, we utilize satellite imagery from the Geostationary Operational Environmental Satellite-16 (GOES-16)~\cite{tan2019goes}. The Advanced Baseline Imager onboard GOES-16 offers high spatial and temporal resolution across multiple spectral bands, spanning wavelengths from visible to infrared. Four channels are utilized as a part of the input data, covering the detection of vegetation, low clouds and fog, and lower tropospheric water vapor. 
Finally, we also include radar data from the Multi-Radar/Multi-Sensor (MRMS) system~\cite{zhang2016multi}. We use the seamless hybrid scan reflectivity as the best estimate of near-surface precipitation. These satellite and radar channels can provide information not present in station observations, such as upper air conditions and information over areas not covered by weather stations.

For additional context, we also add time encodings and topography into the model inputs. The time encodings comprise four channels of sinusoidal signals: \textit{hour\_sin} and \textit{hour\_cos} derived from the time of day, and \textit{month\_sin} and \textit{month\_cos} derived from the month of the year, thereby capturing seasonal patterns. Finally, we incorporated 18 constant topographical variables into the inputs to help the model distinguish surface features across different regions and climates. The topography is derived from multiple data sources such as ERA5, Digital Evaluation Models (DEMs)~\cite{balasubramanian2017digital} and NASA's Ocean Biology Processing Group (OBPG)~\cite{cawse2021nasa}.
A list of all input data sources and their characteristics is provided in Table~\ref{tbl:input_data}, and the meanings of these variables are described in Table~\ref{tb:deschannels}.

\begin{table}[htbp]
\caption{\textbf{Data sources and statistics used in the training phase of \ours{}.}}
\label{tbl:input_data}
\centering
\begin{tabular}{c|c|c|c}
\toprule
\textbf{Type} & \textbf{Source} & \textbf{\#Channels} & \textbf{\#Time Slices} \\
\midrule
\multirow{5}{*}{Input} & GOES Satellites & 4& 6 \\
 & MRMS Radar& 1& 6\\
& Weather Stations & 9& 6 \\
& Time Encodings & 4& 6\\
& Topography & 18& 1 \\
\midrule
Target & RTMA & 6& 18\\
\bottomrule
\end{tabular}
\end{table}

The labels used to train \ours{} are derived from the Real-Time Mesoscale Analysis (RTMA)~\cite{Maunel_2011_RTMA}. During evaluation, we assess the model's performance using both RTMA and station data. For station-level evaluation, we specifically select about 1,600 high-quality stations (approximately 8\% of total stations) as labels. Data from January 1, 2018, to December 31, 2022 are used for training. The validation dataset includes data from the 6th to the 8th day of each month in 2023, while the test dataset includes data from the 1st to the 3rd day of each month in 2023. To assess the generalization capability of our model, \ours{}, we randomly selected a set of hold-out stations (about 250) from those high-quality stations, excluded from the training process. For the station-wise evaluation in Fig.~\ref{fig:OMG-HD+_RTMA}b, Q is derived using relative humidity (RH), T, and SP due to its unavailability from the stations. Similarly, for the hold-out evaluation in Fig.~\ref{fig:OMG-HD+_RTMA}c, the same conversion process is applied; however, SP from RTMA is used as a substitute for station SP. Given the extremely limited availability of stations with valid SP observations, RTMA SP is utilized to ensure sufficient data coverage for the analysis.


All gridded data (including those for \ours{} and other NWP baselines) are interpolated to a $512 \times 1280$ grid with a spatial resolution of 0.05 degree, which approximates the resolution of RTMA. For station data, we map the stations to their nearest grid cell, averaging when more than one station is mapped to one cell. Then inputs and labels are normalized to the range of [-1, 1], while grid points without any available observations are filled with a value of 0.


\subsection{Architecture}\label{model_arch}

\subsubsection{Main blocks}

\ours{} consists of two parts: the Assimilating Block, which transforms the input observations into an initial state, and the Forecasting Block, which makes the forward predictions. This dual-block architecture allows us to use optimal architectures for each of the tasks: The Assimilating Block leverages the Swin Transformer block~\cite{liu2022swin}, which effectively captures localized features with its attention mechanism. This makes it an excellent choice for purposes of data completion and feature extraction. The Forecasting Block incorporates the core architecture of AFNO, which utilizes Fourier transforms to efficiently model meteorological dependencies across different scales.




\noindent\textbf{Assimilating block.} 
The Assimilating Block is built to convert raw observational data into structured embeddings using the modified Swin Transformer V2~\cite{liu2022swin} (SwinT V2). As shown in Fig.~\ref{fig:framework}b, similar to SwinT V2, the Assimilating Block starts with a 3D PatchEmbed layer for tokenization. In this layer, information across all time frames is fused, removing time dimension. This layer is followed by a feature extraction process consisting of SwinT blocks. By leveraging the shifted-window attention mechanism, sparse observations from the surrounding grid points are gathered to form a robust and accurate estimate of the current conditions. Notably, our design differs from the conventional SwinT structure by employing only the first stage (a single Swin Transformer block)~\cite{liu2022swin}, removing patch merging layers. This choice is based on the need to focus on finer details rather than coarse, global features when translating sparse input data into dense outputs. A skip connection is additionally added at the end of the SwinT blocks from the patch embeddings. Finally, the patches, including the time dimension, are recovered using a 3D Unpatchify layer.

As shown in Fig.~\ref{fig:training_loss}, the output of the Assimilating Block includes two parts. The first part consists of 6 channels, which correspond to the 6 variables in the RTMA labels. This part is used to compute the loss with the labels, thereby forcing the model to learn the mapping from sparse observations to dense representations. The second part consists of 20 channels, which do not participate in the loss computation. Instead, these 20 channels are designed to encode and propagate implicit information, which possibly includes upper-air dynamics. Together, these 26 channels serve as input to the Forecasting Block, allowing the model to generate a complete meteorological representation to use for prediction.

\noindent\textbf{Forecasting block.} 
The Forecasting Block uses the learned initial states to make accurate autoregressive predictions. As depicted in Fig.~\ref{fig:framework}b, unlike the Assimilating Block, this block starts with a 2D patch embedding layer where the same patching parameters are applied to each time frame, preserving temporal relationships to facilitate time-series forecasting. The time and feature dimensions are then combined for the subsequent Adaptive Fourier Neural Operator (AFNO)~\cite{guibas2021adaptive} layers. After the AFNO layers, a 2D Unpatchify layer is attached to recover the patches. The output of this block has the same format as that of the Assimilating Block, with only the first 6 channels contributing to the loss computation.

\subsubsection{Training}

\begin{figure}[htbp]
    \centering
    \includegraphics[width=1\linewidth]{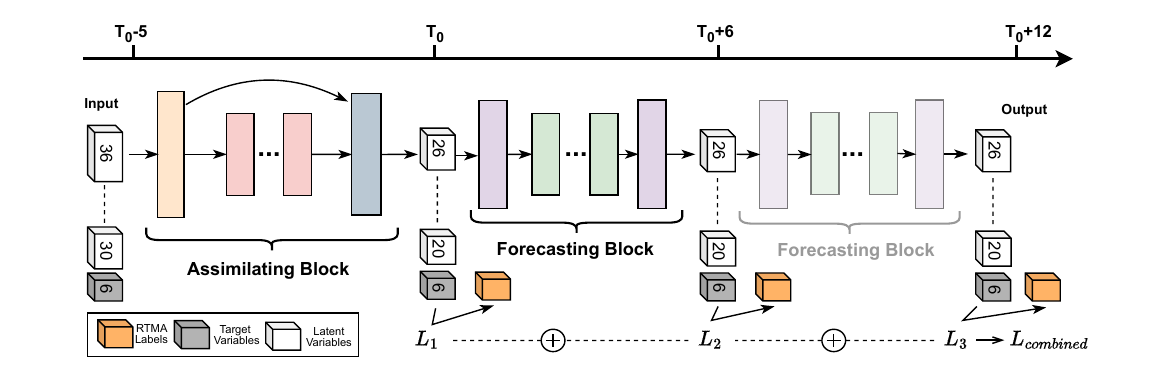}
    \caption{\textbf{Input, output and loss calculation in the training procedure.}} 
    \label{fig:training_loss}
\end{figure}

Each of the 36 total variables is represented as a two-dimensional field with $512 \times 1280$ pixels. These variables are stacked together along the time dimension, forming input tensors of shape $(6,512,1280,36)$. These data tensors are fed into the Assimilating Block in batches to start the end-to-end training process, diagramed in Fig.~\ref{fig:training_loss}. We denote the modeled variables collectively with the tensor $X(T_{start}, T_{end})$ to represent the close interval from the start time to the end time. The RTMA dataset serves as the reference label, denoted by $X_{\text{RTMA}}(T_{start}, T_{end})$.

During training, the model is optimized by a loss function consisting of three parts. The first loss, $L_1$, measures how well the first 6 channels predicted by the Assimilating Block match the gridded observations.
Using RTMA labels from the preceding 6 hours, we can derive $L_1 = f_L\left(X(T_0-5,...,T_0), X_{\text{RTMA}}(T_0-5,...,T_0)\right)$, where, for simplicity, we denote the loss function as $f_L(\cdot)$. 
In a similar manner, $L_2$ is evaluated against the RTMA labels over the next 6-hour sequence using the output of the Forecasting Block, expressed as $L_2 = f_L\left(X(T_0+1,...,T_0+6), X_{\text{RTMA}}(T_0+1,...,T_0+6)\right)$. Finally, in the second autoregressive step, the previous step's output from the Forecasting Block is passed again through the forecaster to compute the third loss, $L_3 = f_L\left(X(T_0+7,...,T_0+12), X_{\text{RTMA}}(T_0+7,...,T_0+12)\right)$. The combined loss, $L_{\text{combined}} = L_1 + L_2 + L_3$, is an equally-weighted sum of all the losses.

The loss function $f_L(\cdot)$, used consistently across all three parts of training in \ours{}, is defined as a weighted sum of Mean Squared Error (MSE) and Mean Absolute Error (MAE). MSE penalizes larger errors more heavily, making it effective at reducing large outliers in predictions. MAE, on the other hand, is more robust to small errors and less sensitive to outliers. Adding these two losses together improves the trained model's accuracy. In our implementation, the MSE and MAE are weighted equally.

During the training phase, we use the Adam~\cite{kingma2014adam} optimizer with a batch size of 64, a weight decay of \(1\times10^{-4}\) and an Epsilon value of \(1\times10^{-8}\). The whole training process runs for 194 epochs (around 50k steps), with the learning rate decaying from \(2 \times 10^{-4}\) to \(5 \times 10^{-7}\). Training takes about two weeks of wall-clock time with 32 AMD Instinct MI250X GPUs using the DeepSpeed~\cite{rasley2020deepspeed} platform.

\section{Additional results}
\label{app:additional}
We supplement several additional experimental results in this section. 

\subsection{Spatial distribution of RMSE for additional variables}
\begin{figure}[htbp]
    \centering
    \includegraphics[width=0.6\linewidth]{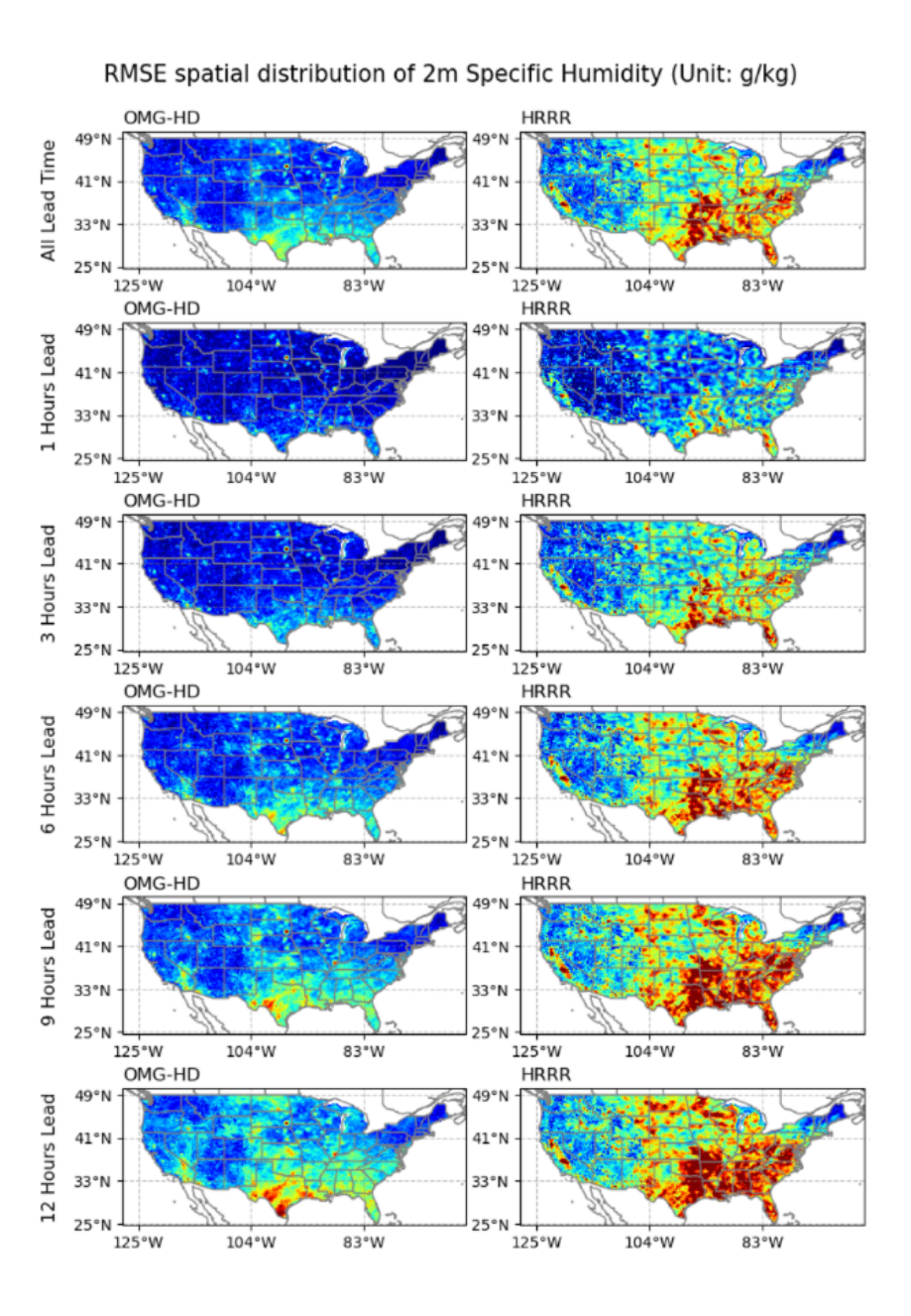}
    \caption{\textbf{RMSE spatial distribution of variable Q with varying lead times.} As our ECMWF and GFS data lack the Q variable, their corresponding heatmaps are absent. In comparison with HRRR, \ours{} generally demonstrates superior performance across all lead times.}
    \label{fig:hum_vis}
\end{figure}

\begin{figure}[htbp]
    \centering
    \includegraphics[width=0.8\linewidth]{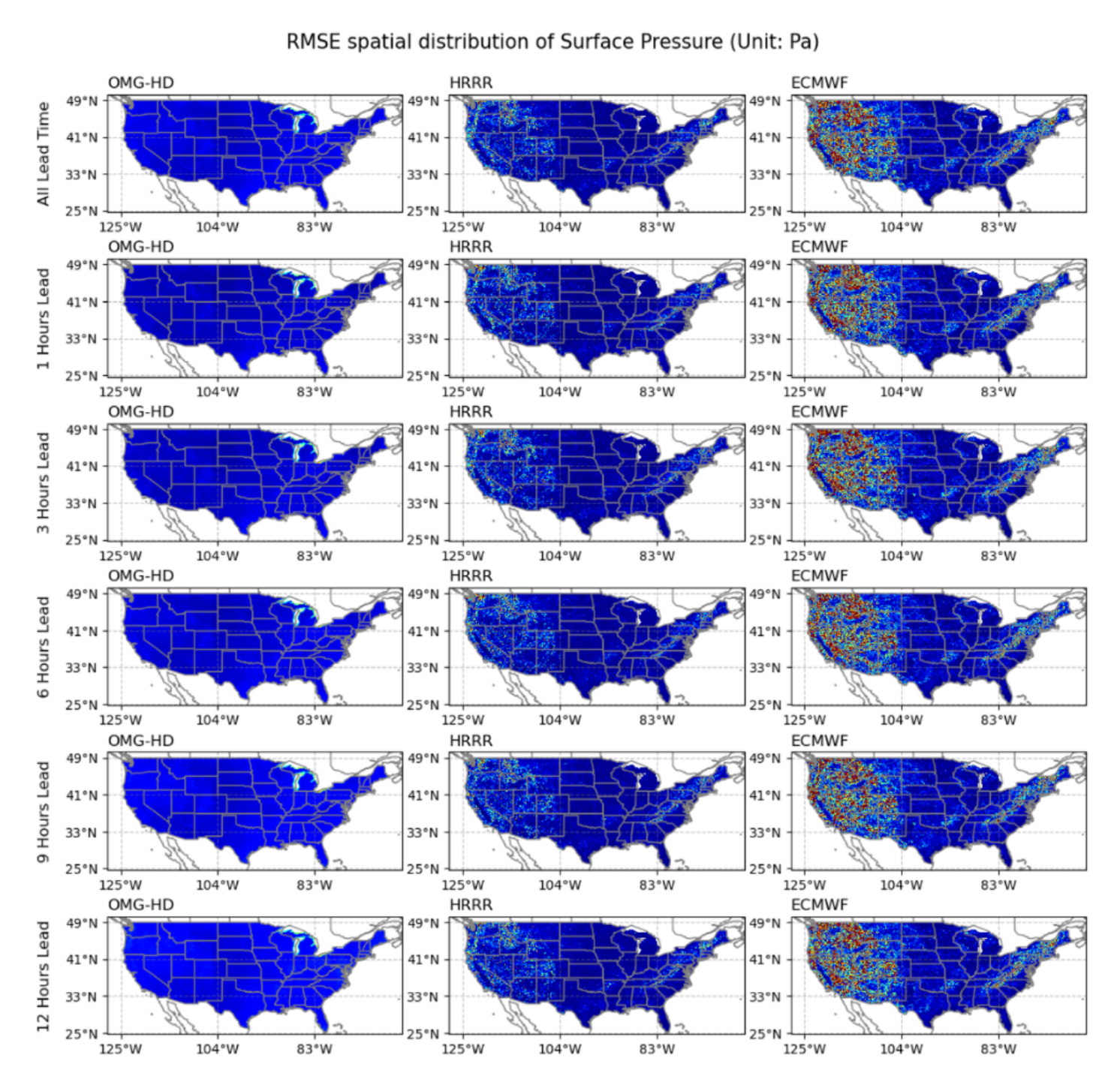}
    \caption{\textbf{RMSE spatial distribution of variable SP with varying lead times.} As our GFS data lacks the SP variable, its corresponding heatmap is absent. }
    \label{fig:pres_vis}
\end{figure}

Fig.~\ref{fig:hum_vis} visualizes the performance of forecasting accuracy on variable Q across the CONUS. \ours{} shows significantly lower error levels, especially for short lead times (1-3 hours), with a deep blue RMSE heatmap across the CONUS region, and maintains better accuracy for long lead times (9-12 hours) compared to HRRR. In the case of variable SP (Fig. \ref{fig:pres_vis}), \ours{} provides superior accuracy over HRRR and IFS-HRES at all lead times, with clear distinctions in lower RMSE. 

\subsection{Supplementary results for the heatwave case}
\begin{figure}[htbp]
    \centering
    \includegraphics[width=1\linewidth]{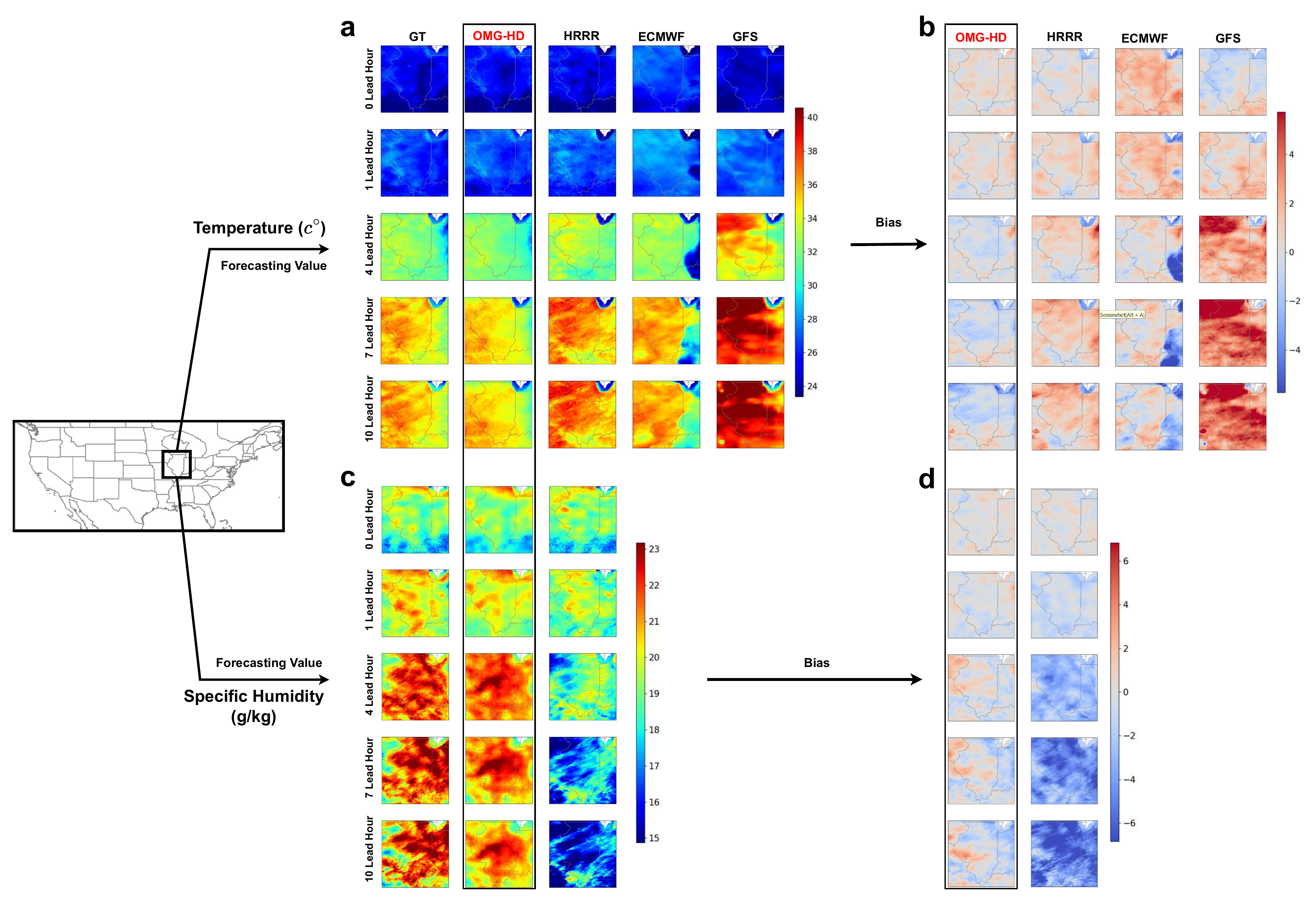}
    \caption{\textbf{\ours{} also excels in predicting temperature and specific humidity during the heatwave case.} Comparison of prediction accuracy for different models during the heatwave event of August 24, 2023 around Chicago, USA. Forecast values and their biases from ground truth over the selected area, where the columns correspond to different models, and rows show different lead times. \ours{} consistently exhibits the lowest bias across all lead times. \ours{} produces the best prediction of both temperature and specific humidity compared to baseline models.
    }
    \label{fig:case_t_q}
\end{figure}

Complementing Section~\ref{cases}, we also show results for temperature (T) and specific humidity (Q) in Fig.~\ref{fig:case_t_q}. For Q, only the HRRR forecast was readily available. For both variables, \ours{} demonstrates the lowest bias among all baseline models. Additionally, in Fig.~\ref{fig:case_t_q}c and d, the performance gap is particularly significant for specific humidity, where \ours{} successfully captures the humidity increase, whereas the HRRR model fails. The improved forecast accuracy for specific humidity also explains why, as shown in Fig.~\ref{fig:heatwave}, \ours{} provides better heat index forecasts compared to the baseline models.

\begin{table}[ht]
\centering
\resizebox{\textwidth}{!}{ 
\begin{tabular}{lll}
\toprule
\textbf{Variable} & \textbf{Unit} & \textbf{Definition} \\
\midrule
\multicolumn{3}{l}{\textbf{Input Variables}} \\
sta\_t & °C & Station temperature at 2 meters \\
sta\_q & g/kg & Station specific humidity \\
sta\_u10 & m/s & Station u component of wind at 10 meters \\
sta\_v10 & m/s & Station v component of wind at 10 meters \\
sta\_rh & \% & Station relative humidity \\
sta\_ws & m/s & Station wind speed at 10 meters \\
sta\_wd & degrees & Station wind direction \\
sta\_p & hPa & Station surface pressure \\
sta\_msl & hPa & Station mean sea level pressure \\
goes1$\sim$4 & - & Satellite channels measuring 0.640, 3.9, 7.4, and 11.20 $\mu$m wavelengths \\
SHSR & dBZ & Seamless hybrid scan reflectivity \\
d2c & km & Distance from coast GMT intermediate \\
lc & - & Land cover class defined in LCCS \\
dl & m & Lake depth \\
cl & - & Lake cover \\
slt & - & Soil type \\
cvh & - & Low vegetation cover \\
cvl & - & High vegetation cover \\
tvh & - & Type of high vegetation \\
tvl & - & Type of low vegetation \\
anor & radians & Angle of sub-grid scale orography \\
isor & - & Anisotropy of sub-grid scale orography \\
z & $\text{m}^2/\text{s}^2$ & Orography / Geopotential \\
lsm & - & Land-sea mask \\
slor & - & Slope of sub-grid scale orography \\
sdfor & m & Standard deviation of filtered sub-grid orography \\
sdor & - & Standard deviation of orography \\
longitude & degrees & Geographical longitude coordinate \\
latitude & degrees & Geographical latitude coordinate \\
hour\_sin & - & Sine transformation for capturing hourly cycles \\
hour\_cos & - & Cosine transformation for capturing hourly cycles \\
month\_sin & - & Sine transformation for capturing monthly cycles \\
month\_cos & - & Cosine transformation for capturing monthly cycles \\
\midrule
\multicolumn{3}{l}{\textbf{Output Variables}} \\
T & °C & Temperature at 2 meters \\
Q & g/kg & Specific humidity \\
WS & m/s & Wind speed at 10 meters \\
SP & hPa & Surface pressure \\
u10 & m/s & U component of wind at 10 meters \\
v10 & m/s & V component of wind at 10 meters \\
\bottomrule
\end{tabular}
} 
\caption{Definition of each variable in the input and output channels.}
\label{tb:deschannels}
\end{table}

\section{Acknowledgments}
The authors express their gratitude to NOAA for providing publicly accessible RTMA analysis data and GOES satellite data.
The authors would like to thank Synoptic Data PBC (accessible at \url{https://synopticdata.com/}) for aggregating station observations and providing the Mesonet API for us to download those data, which are crucial in building an AI-based framework like \ours{} for weather forecasting.


\end{document}